\documentstyle[aps,twocolumn,pra,epsf,eqsecnum]{revtex}
\begin{document} 

\title{An optical trap for collisional studies on cold fermionic potassium}

\author{G. Roati$^\ddag$, W. Jastrzebski\thanks{also Institute of
    Physics, Polish Academy of Science, 02~668 Waszawa, Poland}, A.
  Simoni, G. Modugno\thanks{modugno@lens.unifi.it; http://www.lens.unifi.it}, and M. Inguscio} \address{INFM, LENS, and
  Dipartimento di Fisica, Universit\`a di Firenze,
  Largo E. Fermi 2, I-50125 Firenze, Italy.\\
  $^\ddag$Dipartimento di Fisica, Universit\`a di Trento, I-38050 Povo
  (Tn), Italy}

\date{\today}
\draft
\maketitle

\begin{abstract}
  We report on trapping of fermionic $^{40}$K atoms in a red-detuned
  standing-wave optical trap, loaded from a magneto-optical trap.
  Typically, 10$^{6}$ atoms are loaded at a density of
  10$^{12}$~cm$^{-3}$ and a temperature of 65~$\mu$K, and trapped for
  more than 1~s. The optical trap appears to be the proper environment
  for performing collisional measurements on the cold atomic sample.
  In particular we measure the elastic collisional rate by detecting
  the rethermalization following an intentional parametric heating of
  the atomic sample. We also measure the inelastic two-body
  collisional rates for unpolarized atoms in the ground hyperfine
  states, through detection of trap losses.
\end{abstract}

\section{Introduction}
The fermionic isotope of potassium $^{40}$K is among the most
interesting atomic species for the experimental study of quantum
degenerate gases, as early discussed in \cite{cataliotti}.  Recently,
the possibility of cooling a sample of fermionic $^{40}$K atoms to the
regime of quantum degeneracy has been demonstrated \cite{demarco99},
by means of evaporative cooling in a magnetic trap. This is the first
step towards further studies on a dilute Fermi system, including the
possibility of Cooper pairing. As for this prospect, the presence of
Feshbach resonances in the scattering length of selected Zeeman
substates of $^{40}$K \cite{bohn00}, induced by a magnetic field,
could be exploited to bring the required temperatures for a BCS-like
phase transition to a range experimentally achievable.\par

Optical traps appear to be very useful tools for investigating such
collisional properties of atoms at low temperature, since they can
hold high-density samples in any spin state, and an homogeneous
magnetic field can be applied without modification of the trapping
potential \cite{vuletic99}. For the particular case of $^{40}$K, the
most interesting Feshbach resonance has been predicted for two spin
states which are not magnetically trappable, and therefore an
optical trap would be necessary \cite{bohn00}.\par

In this paper we report for the first time on optical trapping of a
sample of $^{40}$K, precooled in a magneto-optical trap to sub-Doppler
temperatures \cite{modugno99}. By using a compression phase followed
by optical molasses cooling we obtain trapped samples at relatively
low temperature (65~$\mu$K) and high density
(10$^{12}$~cm$^{-3}$). In this temperature and density
regime, our optical trap appears to be an interesting environment to
investigate the collisional properties of $^{40}$K, possibly including
the predicted Feshbach resonances. We are indeed able to perform
elastic collisional measurements on unpolarized sample of atoms in the
ground F=9/2 state, and inelastic measurements on samples in the
F=9/2 and F=7/2 hyperfine substates, which are both held in the
optical trap. A comparison with theoretical predictions is made in both
cases, and a good agreement is found.

\section{Loading and characterization of the optical trap}

\begin{figure}
\begin{center}
\leavevmode\epsfxsize=7cm
    \epsfbox{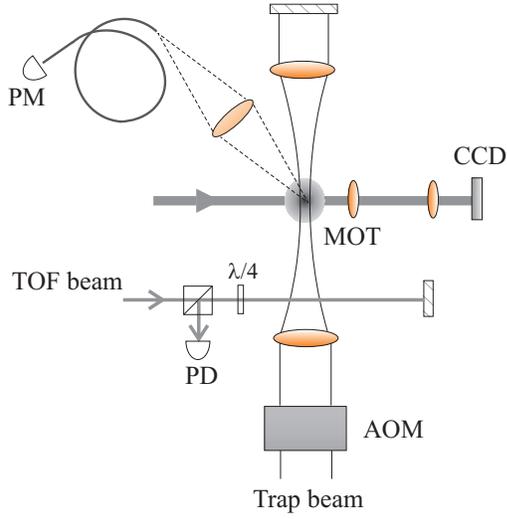}
\end{center}
\caption{Schematics of the experimental apparatus. PD: photodiode; PM:
    photomultiplier.}
\label{apparatus}
\end{figure}

The optical trap is realized with linearly polarized light from a
single-mode Ti:Sa laser, detuned to the red of both the D$_{1}$ and
D$_{2}$ transitions of potassium, respectively at 769.9~nm, and
766.7~nm. The laser radiation is arranged in a vertical standing-wave
configuration, by retroreflecting the beam, achieving a 1D optical
lattice with the strongest confinement against gravity. The laser beam
is weakly focused within a two-lens telescope to a waist-size
$w_0$=90~$\mu$m, with a Rayleigh length $z_R$=3~cm; the effective
running wave power at the waist position is P=600~mW. A schematic
of the experimental apparatus is shown in Fig.~\ref{apparatus}.
\par The
use of a tunable Ti:Sa laser allows us to change in a continuous way
the trap detuning from a few nanometers to about 40~nm, allowing a
variety of trap conditions. For a typical wavelength of the trapping
laser $\lambda_{t}$=787~nm, the calculated trap depth\cite{grimm} for
the ground state is $U_{0}$=300~$\mu$K (6.2~MHz) and the
scattering rate is $\Gamma_{sc}$=30~s$^{-1}$. Since the trapping
light is very far detuned from any transition starting from the
excited 4P states, the light shift for both the excited states is
positive, and it is approximately half of the trap depth.
\par
The vibrational frequencies of each lattice site are given, in the
harmonic approximation, by
\begin{eqnarray}
\omega_{A}&=&2\pi\sqrt{2U_{0}/M\lambda_{t}^{2}}\\\nonumber
\omega_{R}&=&\sqrt{4U_{0}/Mw_{0}^{2}}
\end{eqnarray}
in the axial and radial directions, respectively. Here $M$ is the mass
of potassium. The expected values of the radial and axial trap
frequencies for the conditions above are respectively
$2\pi\times$870~Hz and $2\pi\times$450~kHz.\par The power of the
trapping laser can be controlled with a fast AOM (with a fall time of
300~ns); a polarizer cube after the AOM provides the required
linear polarization. The pump laser is a low-noise doubled Nd:YVO, and
no active stabilization or tuning of the Ti:Sa laser frequency is
used, to avoid amplitude noise which can possibly heat up the trapped
atoms \cite{heating}.\par

We load the optical trap directly from a vapor cell magneto-optical
trap (MOT) of $^{40}$K as described in \cite{modugno99}. We note the
typical cooling and repumping light intensities of 36~mW/cm$^2$ and
4~mW/cm$^2$ respectively, and the detuning of the cooling light of
-20~MHz from the cycling F=9/2~$\rightarrow$~F$^\prime$=11/2 transition (the
natural linewidth and saturation intensities of $^{40}$K are
$\Gamma$=2$\pi\times$6.2~MHz and $I_s$=1.8~mW/cm$^2$). In the MOT we
collect in 4~s about $5 \times 10^{7}$ atoms at a peak density of
$10^{10}$~cm$^{-3}$ and at a temperature of 60~$\mu$K. The dipole trap
beam, which is passing approximately through the center of the MOT, is left
on during the MOT loading, with no detectable effect on the atom
number or temperature.
\par To load the optical lattice from the MOT we use the following
procedure: the trapped atoms are compressed for 30~ms by ramping the
quadrupole gradient of the MOT from the normal value of 18~Gauss/cm up
to 44~Gauss/cm, while the intensity of the cooling and repumping light
are abruptly reduced to 20~mW/cm$^2$ and 0.06~mW/cm$^2$ respectively.
Moreover, the detuning of the cooling light is reduced to -6~MHz to
improve the compression effect. The MOT quadrupole field is then
switched off in 1~ms, and the detuning of the cooling light is
increased to -20~MHz for other 2~ms, providing the usual optical
molasses cooling. Then the MOT beams are turned off by switching off
the AOMs in a few $\mu$s. The repumping beams are turned off 1~ms
later than the cooling ones, to pump the atoms in the lower hyperfine
ground state, F=9/2. For a complete shielding of the optical trap from
the resonant stray light of the MOT beams, which could cause undesired
optical excitation and pumping, we also block the light with a
mechanical shutter within 3~ms from the switch-off of the AOMs. Using
this procedure, about 5\% of the atoms in the MOT can be
loaded in the optical trap, corresponding to a few 10$^6$ atoms.\par

\begin{figure}
\begin{center}
\leavevmode\epsfxsize=9cm
    \epsfbox{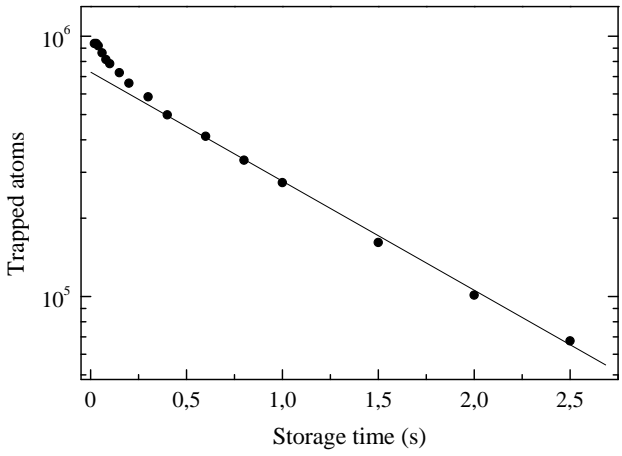}
\end{center}
\caption{Decay of the number of atoms in the optical trap, in the
    ground state F=9/2. The fit with an exponential decay yields a time
    constant of 1.4~s.}
\label{decayg}
\end{figure}

The atoms in the optical trap are probed by turning on again the MOT
beams at half the normal intensity and in resonance with the cycling
transition, while the repumper is in resonance with the
F=7/2~$\rightarrow$~F$^\prime$=9/2. The fluorescence is detected by a
photomultiplier with the help of a lens and of an optical fiber with a
core diameter of approximately 0.5~mm, which provides good spatial
selectivity to discard the light scattered by the windows of the
vacuum cell, and allows detection of as few as 10$^3$ atoms. A typical
measurement of the lifetime of the optical trap is shown in
Fig.~\ref{decayg}. After about 0.3~s of storage, the decay curve of
the atoms pumped in the ground F=9/2 state is a single exponential,
with a 1/e lifetime of $\tau$=1.4(3)~s. We expect the main source of
loss in this regime to be collisions with background gas, the trap
lifetime being roughly one third of the MOT loading time. The faster
decay at shorter times could possibly be due to losses connected to an
evaporative cooling process, which we will consider below.\par

\begin{figure}
\begin{center}
\leavevmode\epsfxsize=7cm
    \epsfbox{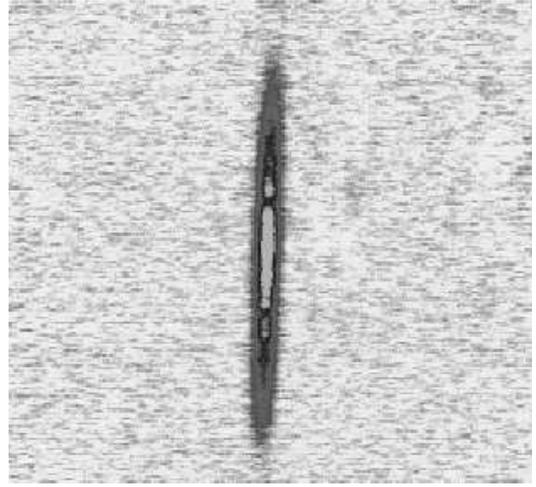}
\end{center}
\caption{Image of the atoms in the optical trap acquired in absorption
    with a CCD camera.The radial (horizontal) FWHM is about 70~$\mu$m, while
    about 1200 individual traps are piled vertically within 500~$\mu$m.}
\label{ccd}
\end{figure}

Alternatively, the density and the absolute number of trapped atoms
can be measured by absorption imaging, using a pulse of light derived
from the MOT laser lasting 200~$\mu$s, at half the saturation
intensity and in resonance with the cycling transition. To compensate
for the light shift due to the optical trap, the frequency is blue
shifted by approximately 1.5$U_0$ from the unperturbed resonance. The
beam is passing through the optical trap in the horizontal plane and
is imaged on a CCD camera with a two-lens system. The distribution of
atoms in the optical trap, as shown in Fig.~\ref{ccd}, is gaussian in
the radial direction, with a typical FWHM of 70~$\mu$m. Also the axial
distribution is generally gaussian, and the typical FWHM of 500~$\mu$m
indicates that approximately 1200 lattice sites, spaced by 395~nm, are
occupied.  The mean number density is determined by averaging the
optical density over the trap section and scaling by the ratio of the
lattice spacing $\lambda_t$ to the the mean axial extension $d_A$ of
an individual lattice site. Such mean axial extension is estimated
from the temperature and axial trap frequency as
\begin{equation}
d_A=\omega_A^{-1}\sqrt{2\pi k_B T/M}\,,
\end{equation}
and is of the order of 100~nm. The typical mean density of atoms
after 100~ms of storage is 7$\times$10$^{11}$~cm$^{-3}$, and
decays proportionally to the number of atoms for increasing storage times.
\par

\begin{figure}
\begin{center}
\leavevmode\epsfxsize=9cm
    \epsfbox{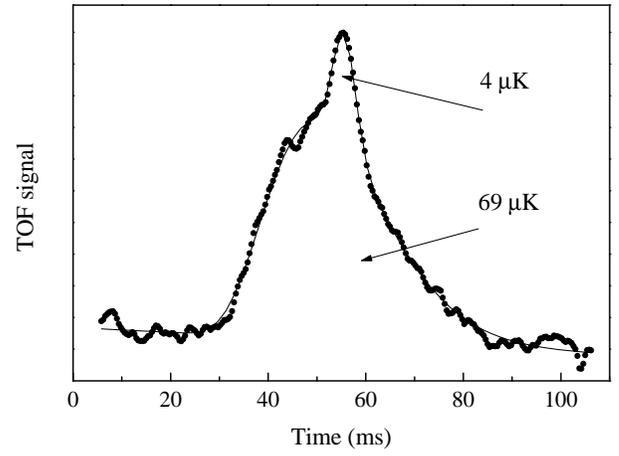}
\end{center}
\caption{Typical TOF signal from the atoms in the optical trap. The
  narrow peak is due to atoms which have been adiabatically cooled
  during the release from the trap. The continuos line is the best fit
  with a two-component TOF distribution, which gives the reported
  temperatures.}
\label{tof}
\end{figure}

The radial temperature is measured by detecting the radial extension
of the atomic cloud after 1~ms of expansion,
following a sudden switch-off of the optical trap. Alternatively it is
possible to estimate the temperature directly from the radial
extensions of the atoms in the trap; this estimation is affected by a
larger uncertainty, since it relies on the knowledge of the radial
trap frequency. The axial temperature is extracted from the temporal
width of a time-of-flight (TOF) signal, as detected by the absorption
of a 1-cm wide sheet of light placed 1.5~cm below the trapping
region, in resonance with the cycling transition. The TOF beam,
generated by an extended-cavity diode laser, is doubly passing through
the vacuum cell, and the absorption signal is detected with a
frequency modulation technique. The typical TOF signal, as shown in
Fig.~\ref{tof}, indicates the presence of two different velocity
distributions, corresponding to temperatures around 70~$\mu$K for the
broad peak and around 4~$\mu$K for the narrow one. We think that the
colder component is due to a small fraction of the trapped atoms which
are adiabatically cooled during the release from the optical
trap. Indeed, the AOM which controls the optical trap is capable of
cutting only 95\% of the light in less than 300~ns, while the residual
tail is extinguished in about 5~$\mu$s. Since the latter switch-off
time is longer than $\omega_A^{-1}$, the atoms in the bottom of the
trap are further cooled down. We have verified this conjecture by
intentionally slowing the main switch-off time of the AOM down to
5~$\mu$s, and observing an adiabatic cooling of hotter component to
8~$\mu$K.  We expect to be possible to eliminate such residual
adiabatic cooling effect by using the AOM in a double-pass
configuration, but since this would reduce by about 30\% the available
laser power, we prefer to fit the TOF signal with a two-component
curve to accurately extract the temperature of the majority of the
atoms.\par The temperature measurement described above can be
performed only after a minimum storage time of 80~ms, to allow for the
separation of the TOF signal due to the atoms in the optical trap from
the signal due to the atoms in the MOT which have not been loaded in the
trap. We have verified that after this time interval the axial and
radial temperature are identical within the uncertainty.\par

\begin{figure}
\begin{center}
\leavevmode\epsfxsize=9cm
    \epsfbox{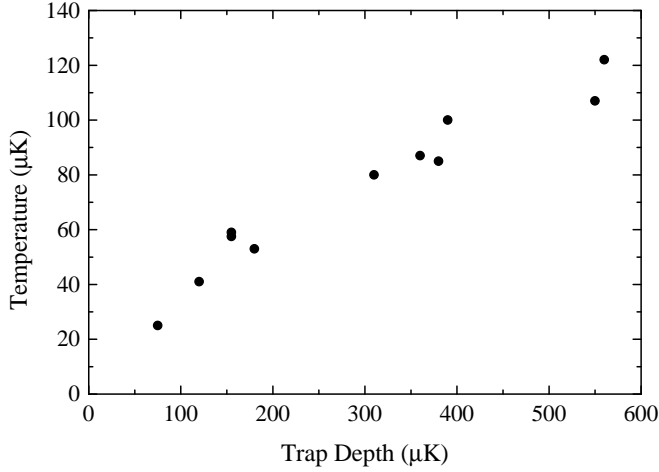}
\end{center}
\caption{Temperature of the trapped atoms after 100~ms of storage as a
    function of the trap depth.}
\label{loading}
\end{figure}

To look for any relationship between the temperatures of the atoms in
the MOT and of those in the optical trap, we have performed a series
of temperature measurements after 100~ms of storage in optical traps
of different depths. The result is shown in Fig.~\ref{loading}: the
temperature appears to scale almost linearly with the trap depth, and
for deep traps we measure a ratio $U_0/k_BT$$\sim$4.5.  Note that for
very deep traps such a temperature can be much higher than the MOT
one, which is about 70~$\mu$K after the compression phase.  This
behavior, which has been reported also in \cite{wiemanpra}, seems to
indicate that the temperature of the atoms in the optical trap is
determined by the process of loading and optical trapping itself, and
not by the original temperature in the MOT. \par We note also that the
number of atoms loaded in the trap is observed to decrease almost
linearly with the trap depth, with the density decreasing even faster
due to the accompanying reduction in the spring constant of the trap.
As a result, the optimal density conditions for performing collisional
measurements are found only at large trap depths, where the
temperature of the trapped atoms is comparable to that of the MOT.  A
detailed description of these observations goes beyond the scope of
this paper, and will therefore be presented elsewhere.\par

We have also observed that the temperature of the atoms in the optical
trap remains almost constant with time, although the calculated
scattering rate would lead to a constant heating with a rate as large
as 100~$\mu$Ks$^{-1}$ for the tightest trap investigated. Also this result
is suggesting the presence of an evaporative cooling effect, which
counteracts the heating due to photon scattering.\par

\begin{figure}
\begin{center}
\leavevmode\epsfxsize=9cm
    \epsfbox{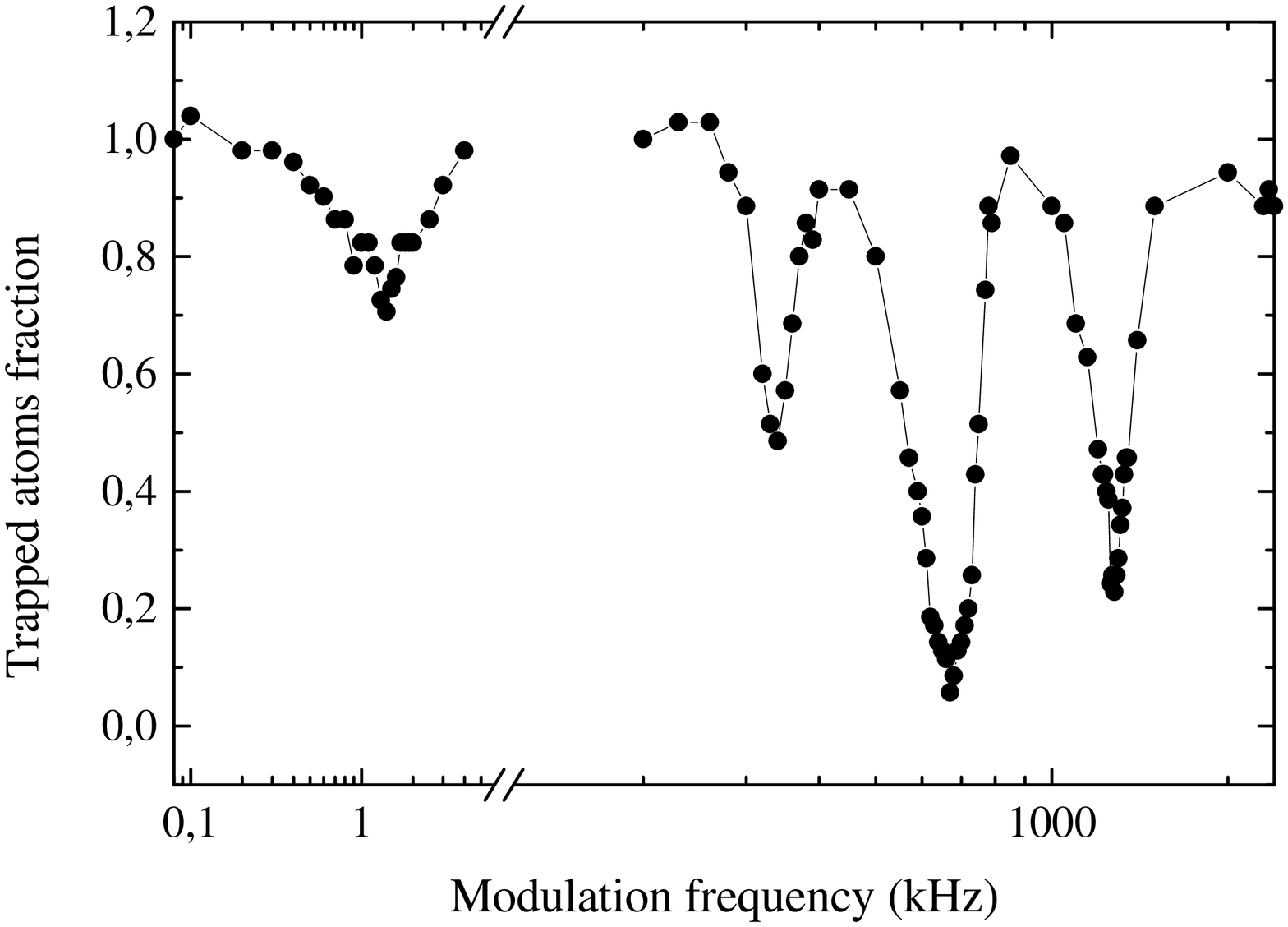}
\end{center}
\caption{Spectrum of the resonances in the parametric heating of the
    atoms in the optical trap. The low-frequency resonance is at twice
    the radial trap frequency, while the high-frequency ones are at
    the fundamental, second and fourth harmonic of the axial trap
    frequency, respectively.}
\label{parametric}
\end{figure}

The effective trap depth and spring constants are measured with the
help of a parametric heating technique \cite{hansch}. If the power of
the trap light is sinusoidally modulated by the AOM, a strong heating
of the atoms, accompanied by trap losses, arises whenever the
modulation frequency matches twice the trap oscillation frequency or a
subharmonic thereof. By monitoring the trap population after
$\sim$100~ms of parametric heating it is therefore possible to measure
both radial and axial trap frequency with an uncertainty of the order
of 10\%. A typical measurement of the trap frequencies is shown in
Fig.~\ref{parametric}, for a trap depth of 180~$\mu$K. Note the
presence of a strong resonance also at four times the axial frequency,
which can be explained only by considering the anharmonicity of the
optical trap in the axial direction, due to its sinusoidal shape
\cite{landau}.
\par

\section{Collisional measurements}
\subsection{Elastic collisions}
\begin{figure}
\begin{center}
\leavevmode\epsfxsize=9cm
    \epsfbox{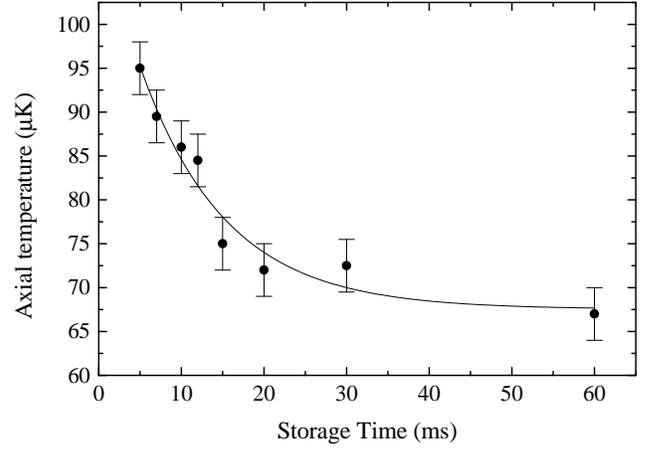}
\end{center}
\caption{Decay of the axial atomic temperature as a consequence of the
    rethermalization with the radial degrees of freedom of the trap,
    mediated by elastic collision. The best fit to an exponential
    decay yields a time constant $\tau$=10(2)~ms.}
\label{decay}
\end{figure}

To characterize the elastic collisions we use a technique which has
already proved useful in this kind of optical trap\cite{vuletic99},
which is based on the detection of the rethermalization of the trapped
atoms after the excitation of one of the degrees of freedom. In
particular, we parametrically excite the axial vibrational mode of the
trap by applying a 2\% modulation to the laser power for about 5~ms,
and measure the subsequent rapid increase of the axial temperature,
followed by an exponential decay due to thermalization of the axial
mode with the radial ones, as mediated by elastic collisions. In order
to excite uniformly the atoms distributed over the vibrational levels
in the anharmonic trap, we sweep the modulation frequency roughly
between $\omega_A$ and 2$\omega_A$ during the 5~ms heating phase.
Indeed, if only a single frequency is used, we observe a large loss of
atoms from the trap, with no larger increase of the temperature, as if
only the population of a few levels were excited. Moreover, modulation
amplitudes larger than 2\% cause quite a large loss of atoms, which
are detected by the TOF beam together with those released from the
trap, therefore altering the temperature measurements. The decay curve
for the conditions $T$=~65$\mu$K, $n$=7.5$\times$10$^{11}$~cm$^{-3}$,
is shown in Fig.~\ref{decay}; each data point is the average of two
series of about 20 TOF measurements, and the original temperature was
67~$\mu$K. The fit to an exponential decay yelds $\tau_{th}$=10(2)~s.
We have checked that the decay time is halved, within the uncertainty,
when the atomic density in the optical trap is halved, to verify that
the rethermalization is actually mediated by elastic collisions, and
not simply by cross-dimensional anharmonic effects.\par

We compare the observed collision rates with the results of a full
close-coupling numerical calculation with the standard Hamiltonian
including hyperfine structure.  The model is parametrized by the long
range dispersion coefficients of the atom-atom interaction potential
and by single channel singlet and triplet scattering lengths.  We
adopt the high precision value $C_6=$3897au from
Ref.\cite{Derevianko} for the van der Waals coefficient, and choose
the range of possible scattering lengths by scaling to $^{40}$K the
results obtained for $^{39}$K from an analysis of the $0_g^-$
\cite{0g-} molecular state photoassociation spectroscopy. Similar but
less restricted values are also presented in \cite{1g} from an
analysis of the $1_g$ spectrum.  We obtain in this way the limits 100
$< a_s< $110a$_0$ and 150 $< a_t < $250a$_0$.  It is consistent with
the experimental errors in the present measurements to fix the singlet
scattering length to say $a_s=105a_0$ and leave only the triplet as a
free parameter.\par From our calculation and from the results of an
analogous experiment performed in a magnetic trap \cite{demarcoprl},
we expect that both $s$- and $p$-wave collisions are contributing to the
cross sections at $T$=65~$\mu$K for $^{40}$K.  Indeed, a $p$-wave shape
resonance close to threshold enhances the $p$-wave contribution at
low temperatures \cite{demarcoprl,bohn99}.  Since our sample is
unpolarized we calculate spherical $s$ and $p$ cross sections
$\sigma$, and estimate the thermalization rate as
\begin{equation}
\frac{1}{\tau_{th}}
= \big(\frac{\sigma_s}{\alpha_s} + \frac{\sigma_p}{\alpha_p} \big)       
nv\,
\end{equation}
where $\alpha$ is the average number of collisions necessary for the
rethermalization, $n$ is the atomic density, and $v$ is the rms
relative velocity between two colliding atoms.  Note that we have
implicitly assumed an identical distribution of all possible spin
states in F=9/2 over the optical trap, since the trap potential is
independent of the magnetic moment of the atoms.  We adopt for the
parameters $\alpha$ the values $\alpha_s=2.5$ and $\alpha_p=4.1$
quoted in \cite{demarcoprl}.  We find the best agreement with the
measured rate at the lowest allowed value of the triplet scattering
length ($a_t$=150$a_0$), {\it i.e.}  when the $p$-wave resonance is
closer to threshold.  In this situation the average thermal rates are
$\sigma_s v$=$6.2 \times 10^{-11}$cm$^3$/s and $\sigma_p v$=$1.75
\times 10 ^{-10}$cm$^3$/s and the characteristic time for
rethermalization for the given density in the trap is then about
$\tau_{th}$=20~ms.  It is interesting to notice that a comparison
between the photoassociation results in \cite{0g-} and \cite{1g} seems
to privilege lower values of $a_t$ in agreement with what we find
here.\par


From the calculated elastic collision rate 1/$\tau_{el} \equiv n
\sigma v$, with $\sigma$ the total cross section, is also possible to
estimate the typical speed of the thermalization due to evaporative
cooling processes. Using the formalism presented in \cite{ketterle},
the ratio of the evaporation to the elastic collision time constant is
\begin{equation}
\tau_{ev}/\tau_{el}\approx\sqrt{2}\,{\rm e}^{\displaystyle \eta}\eta^{-1}\,
\end{equation}
where $\eta$=$U_0$/$k_BT$. From the measured value of $\eta\sim4.5$
(see Fig.~\ref{loading}), the timescale for the thermalization due to
evaporation is therefore expected to be a few hundreds of ms. Although
this process can be neglected with regard to the cross-dimensional
collisional processes reported above, which are proceeding at a much
faster rate, we speculate that it could play an important role in the
observed long term stability of the temperature of the atoms in the
optical trap.
\subsection{Inelastic collisions}
The inelastic two-body collisions between atoms in the F=7/2 state,
and between atoms in the F=9/2 and F=7/2 states, can be detected in a
relatively easy way through trap losses, since the hyperfine energy of
1.285~GHz released under the transition F=7/2$\rightarrow$9/2 is much
larger than the trap depth. We can vary the population of the upper
hyperfine state F=7/2 by means of an optical pumping beam, derived
from the MOT laser, in resonance with the F=9/2$\rightarrow$F$^\prime$=7/2
transition, and controllable in intensity with an AOM.  By shining the
beam on the trapped atoms with an intensity of about 10~$\mu$W/cm$^2$,
and for a time interval variable between 0.1 and 2~ms, we transfer up
to 95\% of the atoms from the ground to the upper state. The relative
populations of the two hyperfine states after the pumping pulse are
measured by monitoring the TOF signal, to which only the atoms in
F=9/2 contribute. We have verified that the typical time scale for
hyperfine optical pumping due to inelastic Raman scattering of the
trap photons is much longer than 1~s, and therefore such effect is
negligible.\par
\begin{figure}
\begin{center}
\leavevmode\epsfxsize=9cm
    \epsfbox{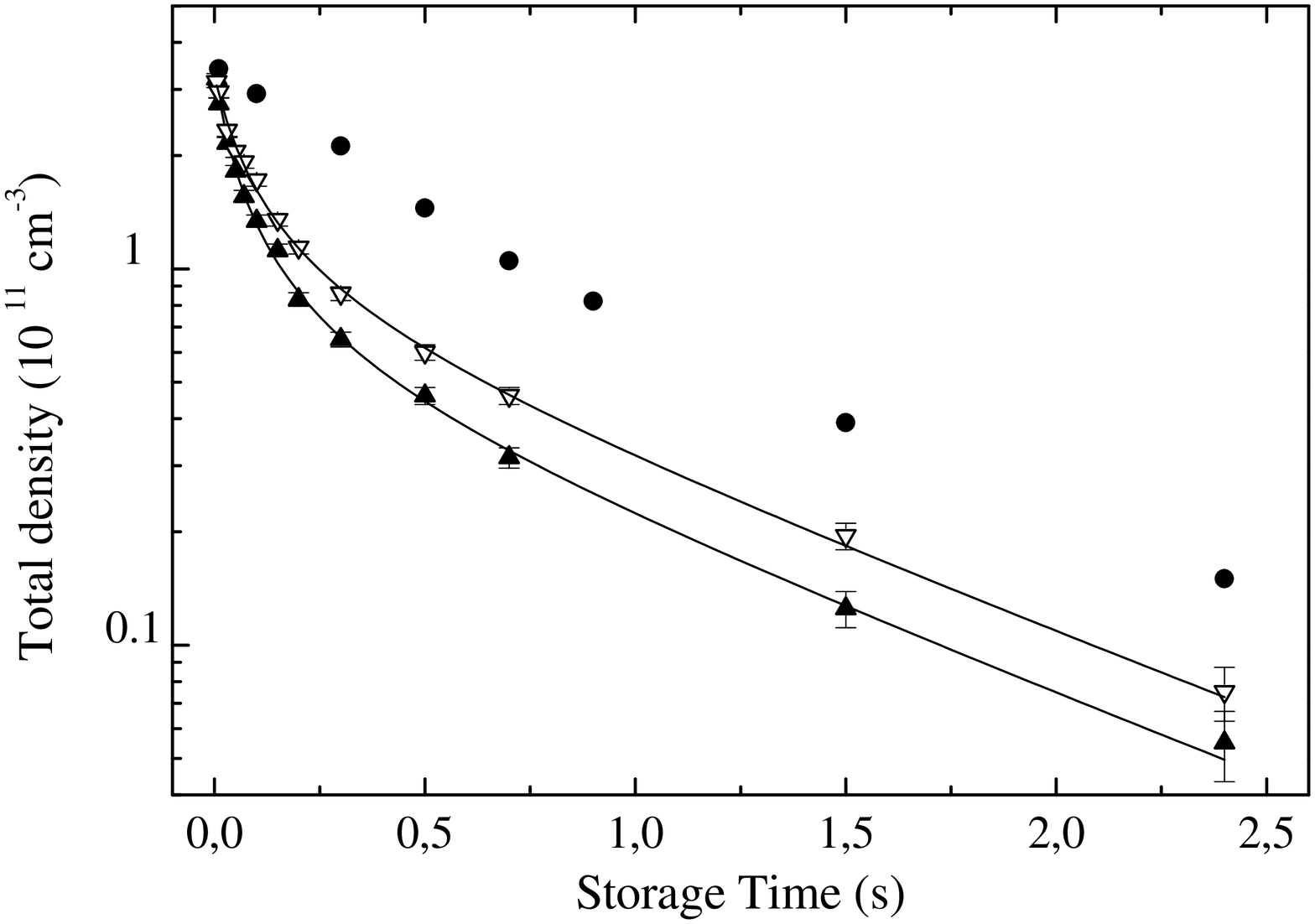}
\end{center}
\caption{Decay of the number of atoms in the optical trap for
  different populations of the upper hyperfine level F=7/2: open
  triangles 87\%; solid triangles 95\%. The lines
  represent the best fit with the solution of the coupled differential
  equation described in the text. The third set of points (circles)
  shows for comparison the decay for the F=9/2 state.}
\label{hyperfine}
\end{figure}

 The optical pumping pulse is applied after about
100~ms from the loading of the dipole trap, to let the
trapped atoms equilibrate, and then the decay of the number of atoms
in the trap is measured with the usual procedure. In
Fig.~\ref{hyperfine} the decay of the trapped atoms is shown for the
cases of 95\% and 85\% of the total population pumped into the F=7/2
state. In this particular measurement the temperature of the atoms was
$T$=50~$\mu$K. We fit the experimental data by assuming a constant volume
regime, which in an harmonic trap is equivalent to the assumption of a
constant temperature. In this regime, the density is proportional to
the number of atoms, and the decay curves are the solutions of the
coupled differential equations for the densities of the populations in
both hyperfine states
\begin{eqnarray}
\dot{n_1}\,&=&\,-\,\gamma n_1\,-
\,G_{1,2}n_{1}n_2\\
\dot{n_2}\,&=&\,-\,\gamma n_1\,-\,2G_{2,2}n_2^2\,-
\,G_{1,2}n_1n_2\,.
\end{eqnarray}
Here the suffixes 1 and 2 refer to the levels F=9/2 and F=7/2,
respectively, $\gamma$ is the rate of linear trap loss, and $G_{i,j}$
is the spherical rate of two-body inelastic collisions between atoms
in states $i$ and $j$ at temperature $T$.  To account for the loss of
both atoms in the pair for each collision event, factors of 2 are
added for collisions of atoms belonging to the same hyperfine level.
\par 
The constant $\gamma$ can be obtained from the tails of the decay
curve, where the binary collision rate is negligible.  The fit yields
$\gamma$=(0.90$\pm$0.05)~s$^{-1}$.  We keep this value fixed for the
fits where we have 95\%, 83\% and 66\% of the atoms in F=7/2.  We
assume a 30\% uncertainty in the atomic density and obtain best
estimates of the decay rates from the individual curves. Next we
combine the three measurements to obtain
$G_{7/2,7/2}$=3.2$\times$10$^{- 11}$~cm$^3$/s and
$G_{9/2,7/2}$=3.8$\times$10$^{-12}$~cm$^3$/s with an uncertainty of
the order of 50\%. From nu\-me\-ri\-cal cal\-cu\-la\-tions we find
thermally averaged spherical rates of
$G_{7/2,7/2}$=4.4$\times$10$^{-11}$~cm$^3$/s and
$G_{9/2,7/2}$=3.5$\times$10$^{-12}$~cm$^3$/s for $T$=50$\mu$K, in
agreement with the measured rate.  We obtain these values at the lower
edge $a_3$=150$a_0$ of the initial confidence range. The agreement
holds until the value $a_3 \simeq 180a_0$, where the loss coefficient
$G_{9/2,7/2}$ drops to values of the order of $G_{9/2,7/2}\simeq
10^{-12}$~cm$^3$/s.  In all this range $p$-wave losses provide a
significant contribution to both $G_{7/2,7/2}$ and $G_{9/2,7/2}$, and
the agreement with the measured values represents therefore further
evidence of the presence of a low energy $p$-wave resonance. It is
interesting to notice that the rate coefficient $G_(9/2,7/2)$ rapidly
increases above $a_3$=180$a_0$, reaching values larger than
10$^{-11}$~cm$^3$/s for $a_3>190a_0$, in disagreement with the measured
value. This seem to rule out the higher range of $a_3$ values, as
already indicated by the large observed rethermalization rate. We also
note that inelastic collisions cannot be responsible for the faster
decay of the atom number at short trapping times (see
Fig.~\ref{decayg}), even assuming a residual 5\% on the F=7/2 state
after the normal loading of the optical trap, due to the relatively
small value of $G_{9/2,7/2}$.

\section{Conclusions}
We have reported on loading of fermionic $^{40}$K atoms in a 1D
optical lattice from a single MOT. The 1.4-s lifetime of the optical
trap is large enough to allow the observation of elastic and inelastic
collisional processes, due to the high density of the atomic sample in
the trap. Indeed, the typical time scale for collisional processes is
a few tens of ms, at a density of 7$\times$10$^{11}$~cm$^{-3}$ and at
a temperature of 65~$\mu$K. Furthermore, optical pumping and heating
effects arising from the scattering of trap photons do not seem to
play a significant role on such time scale.\par The experimental
collisional rates for unpolarized samples we determine are quite
consistent with the theoretical expectations. Further studies of
inelastic and elastic collisions in spin-polarized samples in our
optical trap could lead to a characterization of the collisional
properties of fermionic potassium at low temperature. As already
noted, particularly interesting is the possibility of searching for
the occurrence of $s$-wave Feshbach resonances in the collisions
between Zeeman sublevels of the ground state. For such purpose, a
temperature lower than the present 65~$\mu$K would be preferred, to
reduce the $p$-wave contribution, and to avoid temperature broadening
of the resonance. Although the temperature can be lowered by loading
the atoms in very shallow traps, the achieved density is not large
enough to perform accurate collisional measurements, and therefore
methods for reducing the temperature of a dense sample are needed. In
particular, we note that the axial degree of freedom of our trap is in
the Lamb-Dicke regime ($d_A\ll\lambda$), allowing for the application
of fast degenerate Raman sideband cooling \cite{vuletic98}, which
would allow for an efficient reduction of the temperature without atom
losses. \par

We thank D. Lau for his partecipation to the early phases of the
experiment, and Laboratorio di Fisica della Scuola Normale Superiore
for loaning the Ti:Sa laser for the optical trap. We also acknowledge
contributions by P. Hannaford and N. Poli, and useful discussions
with V. Vuletic and F. S. Cataliotti. One of us (A. S.) gratefully
thanks E. Tiesinga for helpful discussions, and the NIST group for the
ground state collision code.  This work was supported by the European
Community Council (ECC) under the Contract HPRICT1999-00111, and by
MURST under a PRIN Program (Progetto di Ricerca di Interesse
Nazionale). W. J. was funded by the NATO-CNR Senior Guest Fellowship
Programme 1998.

\newpage

\begin{thebibliography}{99}
\bibitem{cataliotti} F. S. Cataliotti, E. A. Cornell, C. Fort,
  M. Inguscio, F. Marin, M. Prevedelli, L. Ricci, and G. M. Tino,
  Phys. Rev. A {\bf 57}, 1136 (1998). 
\bibitem{demarco99} B. DeMarco and D. S. Jin, Science {\bf 258}, 1703
  (1999).  
\bibitem{bohn00}J. Bohn, Phys. Rev. A {\bf 61}, 053409 (2000).
  \bibitem{vuletic99} V.  Vuletic, C. Ching, A.J. Kerman, and S. Chu,
  Phys. Rev.  Lett. {\bf 82}, 1406 (1999).  
\bibitem{modugno99} G.
  Modugno, C.  Benko, P. Hannaford, G. Roati, and M.  Inguscio, Phys.
  Rev. A {\bf 60}, R3373 (1999).  
\bibitem{grimm}See for example: R.
  Grimm, M.  Weidem\"uller, and Y. B.  Ovchinikov, Adv. At. Mol. Opt.
  Phys. {\bf 42} 95 (2000).  
\bibitem{heating} For a discussion of the
  heating of the trapped atoms due to noise in the trap laser, see T.
  A. Savard, K. M. O'Hara, and J. E. Thomas Phys.  Rev. A {\bf 56},
  R1095 (1997).  

\bibitem{wiemanpra}S. J. M. Kuppens, K. L. Corwin, K. W. Miller, T. E.
  Chupp, and C. E. Wieman, Phys. Rev. A {\bf 62} 013496-1 (2000).
\bibitem{hansch} S. Friebel, C. D'Andrea, J. Walz, M. Weitz, and
  T. W. H\"ansch, Phys. Rev. A {\bf 57}, R20 (1998).
\bibitem{landau}For a discussion of the classical case of parametric
  heating of an anharmonic oscillator see L. D. Landau and E. M.
  Lifschitz, {\it Mechanics} (Pergamon, Oxford , 1976), pp. 80-84.

\bibitem{Derevianko} A. Derevianko, W. R. Johnson, M. S. Safronova, and J. 
F. Babb, Phys. Rev. Lett {\bf 82}, 3589 (1999).
\bibitem{0g-} J. P. Burke,Jr., C. H. Greene , J. L. Bohn, H. Wang, P. L. 
Gould, and W. C. Stwalley, Phys. Rev. A {\bf 60}, 4417 (1999).
\bibitem{1g} C. J. Williams, E. Tiesinga, P. S. Julienne, H.Wang, W.C. 
Stwalley, and P. L. Gould, Phys. Rev. A {\bf 60}, 4427 (1999).
\bibitem{demarcoprl} B. DeMarco et al., Phys. Rev. Lett. {\bf 82}, 4208
  (1999).
\bibitem{bohn99}J. Bohn et. al. Phys. Rev. A {\bf 59} 3660 (1999)  
\bibitem{ketterle} W. Ketterle and N. J. van Druten,
  Adv. At. Mol. Opt. Phys. {\bf 37} 181 (1996).
\bibitem{vuletic98} V. Vuletic, C. Chin, A. J.
  Kerman, and S. Chu, Phys. Rev. Lett.{\bf 81}, 5768 (1998).
\end{thebibliography}
\end{document}